\documentclass[preprint]{aastex701}
\usepackage{amsmath,CJK}

\begin{document}
\begin{CJK*}{UTF8}{gbsn}

\title{Prediscovery Activity of New Interstellar Object 3I/ATLAS: A Dynamically-Old Comet?}

\correspondingauthor{Quanzhi Ye}

\author[orcid=0000-0002-4838-7676]{Quanzhi Ye (叶泉志)}
\affiliation{Department of Astronomy, University of Maryland, College Park, MD 20742, USA}
\affiliation{Center for Space Physics, Boston University, 725 Commonwealth Ave, Boston, MA 02215, USA}
\email[show]{qye@umd.edu}

\author[orcid=0000-0002-6702-7676]{Michael S. P. Kelley}
\affiliation{Department of Astronomy, University of Maryland, College Park, MD 20742, USA}
\email{msk@astro.umd.edu}

\author[0000-0001-7225-9271]{Henry H. Hsieh}
\affiliation{Planetary Science Institute, 1700 East Fort Lowell Rd., Suite 106, Tucson, AZ 85719, USA}
\email{hhsieh@psi.edu}

\author[0000-0001-8018-5348]{Eric C. Bellm}
\affiliation{DIRAC Institute, Department of Astronomy, University of Washington, 3910 15th Avenue NE, Seattle, WA 98195, USA}
\email{ecbellm@uw.edu}

\author[0000-0001-9152-6224]{Tracy X. Chen}
\affiliation{IPAC, California Institute of Technology, 1200 E. California Blvd, Pasadena, CA 91125, USA}
\email{xchen@ipac.caltech.edu}

\author[0000-0002-5884-7867]{Richard Dekany}
\affiliation{Caltech Optical Observatories, California Institute of Technology, Pasadena, CA 91125, USA}
\email{rgd@astro.caltech.edu}

\author{Andrew Drake}
\affiliation{Division of Physics, Mathematics, and Astronomy, California Institute of Technology, Pasadena, CA 91125, USA}
\email{ajd@astro.caltech.edu}

\author[0000-0001-5668-3507]{Steven L. Groom}
\affiliation{IPAC, California Institute of Technology, Pasadena, CA 91125, USA}
\email{sgroom@ipac.caltech.edu}

\author[0000-0003-3367-3415]{George Helou}
\affiliation{IPAC, California Institute of Technology, Pasadena, CA 91125, USA}
\email{gxh@ipac.caltech.edu}

\author[0000-0001-5390-8563]{Shrinivas R. Kulkarni}
\affiliation{Division of Physics, Mathematics, and Astronomy, California Institute of Technology, Pasadena, CA 91125, USA}
\email{srk@astro.caltech.edu}

\author[0000-0002-8850-3627]{Thomas A. Prince}
\affiliation{Division of Physics, Mathematics, and Astronomy, California Institute of Technology, Pasadena, CA 91125, USA}
\email{prince@caltech.edu}

\author[0000-0002-0387-370X]{Reed Riddle}
\affiliation{Caltech Optical Observatories, California Institute of Technology, Pasadena, CA 91125, USA}
\email{riddle@caltech.edu}

\begin{abstract}

We report on the prediscovery observations and constraints of the new interstellar comet 3I/2025 N1 (ATLAS), made by the Zwicky Transient Facility (ZTF), for the inbound leg of the comet out to a heliocentric distance of $r_\mathrm{h}=17$~au, or approximately a year before its discovery. We find that 3I/ATLAS has been active inward of a heliocentric distance of at least $r_\mathrm{h}=6.5$~au. The comet followed a brightening rate of $\propto r_\mathrm{h}^{-3.8}$, which is significantly steeper than the only other known interstellar comet 2I/Borisov, and is more consistent with dynamically old long-period comets and short-period comets in the Solar System. By measuring the brightening of the dust coma, we estimate that 3I had a dust production rate of $\dot{M_\mathrm{d}}\sim5~\mathrm{kg~s^{-1}}$ in early May of 2025 ($r_\mathrm{h}\sim6$~au), increasing to $\dot{M_\mathrm{d}}\sim30~\mathrm{kg~s^{-1}}$ towards mid-July 2025 ($r_\mathrm{h}\sim4$~au) assuming 100~\micron{} dust grains, in line with the more recent \textit{Hubble Space Telescope} measurement made at $r_\mathrm{h}=3.8$~au. Comparison with the prediscovery photometry by the \textit{Transiting Exoplanet Survey Satellite} (TESS) suggested that 3I started producing constant dust outflow probably around $r_\mathrm{h}\sim9$~au, coinciding with the turn-on distance of CO$_2$ ice. We also conduct a deep search of 3I/ATLAS with multiple nights of data taken in 2024 when the comet was at $r_\mathrm{h}=13$--$17$~au and conclude that the comet was no brighter than 2--5 magnitudes above the coma or bare-nucleus lightcurves. This suggests that the comet did not exhibit strong outbursts during these periods, consistent with 2I/Borisov as well as most long-period Solar System comets.

\end{abstract}



\section{Introduction} 

Interstellar Objects (ISOs) are exogenous small bodies that traverse the Solar System on orbits gravitationally unbound to the Sun. The discovery of comet 3I/2025 N1 (ATLAS) marks the third identified ISO and the second unambiguous interstellar comet, following the discoveries of 1I/`Oumuamua in 2017 and 2I/Borisov in 2019. 3I/ATLAS was first reported by the Asteroid Terrestrial-impact Last Alert System (ATLAS) survey on UT 2025 July 1, from observations made at the R\'{i}o Hurtado station in Chile \citep{Denneau20253I}. Prediscovery detections by the Zwicky Transient Facility (ZTF) as well as additional prediscovery detections found in ATLAS images, reported within a few hours of the initial detection, suggested an orbit with an eccentricity of $e\sim6$,  indicating an origin beyond the Solar System. The discovery of a second interstellar comet is significant not only for confirming that such objects may be relatively common, but also for enabling comparative studies of composition, activity, and origin between different interstellar comets.

The brightening curve of comets provides valuable clues about the size of the nucleus and its volatile abundances. Here we present an analysis of the prediscovery observations of 3I/ATLAS by ZTF in aim to study the activity and brightness evolution of the comet prior to discovery.

\section{Observations} \label{sec:obs}

ZTF is a wide-field optical survey operating on the 1.2 m Palomar Oschin Schmidt telescope with a dedicated camera covering a 55 deg$^2$ field-of-view \citep{2019PASP..131a8002B, 2019PASP..131g8001G, 2020PASP..132c8001D}.  The camera is a mosaic of 16 6k$\times$6k CCDs with a pixel scale of 1\farcs01.  The camera has a filter exchanger with a set of custom ZTF-$g$, ZTF-$r$, and ZTF-$i$ filters with transmissions tailored to the sky background at Palomar Observatory \citep{2019PASP..131a8002B}, and therefore are slightly different from the similarly-named filter sets used by the Sloan Digital Sky Survey \citep[SDSS;][]{york2000_sdss} and the Pan-STARRS survey \citep{tonry2012_ps1}.  At a survey speed of $3760~\mathrm{deg^2~hr^{-1}}$ (30~s for each exposure), ZTF regularly surveys the visible northern sky every a few nights, making it ideal for the long-term monitoring of small bodies including pre-discovery detections.

Initial ZTF prediscovery detections of 3I were identified a few hours after the initial ATLAS discovery report on UT 2025 July 1 and helped establish its interstellar nature \citep{2025arXiv250702757S}. With the improved orbit, additional ZTF prediscovery detections dating back to UT 2025 May 22 were subsequently identified on individual (unstacked) images.

After the discovery was formally announced with a well-established orbit, we used the {\tt ZChecker} search tool and comet monitoring code \citep{2019ASPC..523..471K} with JPL orbit solution \#16 (the latest JPL solution at the time that the search was conducted, calculated on UT 2025 July 23), and found a total of 187 images taken between UT 2024 June 15 and UT 2025 July 20 that covered the nominal ephemeris position of the comet. All images are standard survey images with 30~s exposure times. {\tt ZChecker} generated filter-separated nightly stacks from these images, resulting in a total of 145 stacked images. Individual ZTF images were pipeline processed by the ZTF Science Data System to remove instrumental signatures, and to derive astrometric and photometric calibrations using the Gaia-DR1 and PS1 catalogs, respectively \citep{2019PASP..131a8003M}.  The calibrations are based on point-source-function (PSF) photometry of the stars in the image.  After calibration, the pipeline subtracts a template image of the static background sky, leaving a ``differenced'' image that primarily contains signatures of variable and moving sources.

\section{Analysis} \label{sec:analysis}

\subsection{Photometry \& Lightcurve}\label{sec:photometry}

To eliminate low-quality data points, we first reviewed the nightly stacks and exclusively used stacks generated from differenced images, as failure in background subtraction often indicates underlying image quality issues. We also excluded images with apparent cloud contamination (limiting magnitude $\ll20$) or bright stars near the comet's position.

\begin{figure*}
\includegraphics[width=\textwidth]{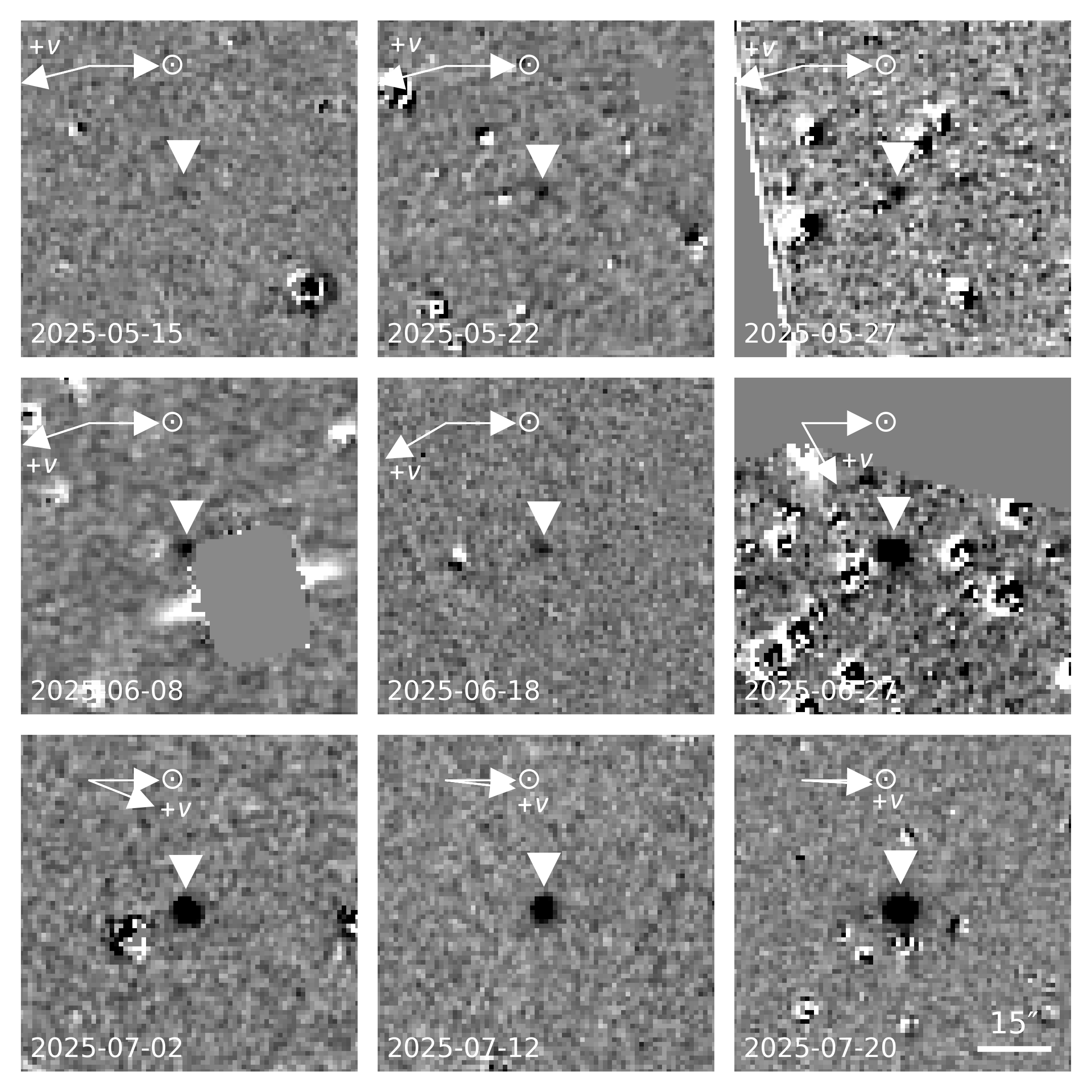}
\caption{A selection of nightly ZTF stacks of 3I from its first detection on UT 2025 May 15 to UT 2025 July 20 (the latest stack used in this work). 3I is marked by a filled triangle in each panel. The panels have North up and East to the left. The arrows in the upper-left of each panel mark the direction to the Sun ($\odot$) and velocity vector ($+v$) of the comet. The images are color-inverted.}
\label{fig:medley}
\end{figure*}

We were able to unambiguously trace 3I on the nightly stacks dating back to UT 2025 May 15, resulting in a total of 41 positive detections (Figure~\ref{fig:medley}). For these detections, we use a $3''$-radius aperture to perform aperture photometry on the comet. The $3''$ aperture radius is chosen considering the seeing of the images (varying between $1\farcs5$ and $4\farcs4$) and crowded fields near the galactic plane in which the comet was traversing.

We then calculated and applied an aperture correction to account for the blurring of the comet by the PSF and derive the ``true'' brightness in the $3''$-radius aperture.  The aperture correction for point sources, calculated by the ZTF pipeline, ranges from --0.60 to --0.08~mag, and the pipeline-generated PSFs can be used to derive aperture corrections for extended sources such as comets. As 3I was only marginally resolved in our data (with the apparent size of the coma no more than a few pixels across), we are unable to meaningfully measure the surface brightness profile of the comet outside of the seeing disk.  Therefore, we assume a nominal $1/\rho$ profile, where $\rho$ is the angular distance to the nucleus.  A $1/\rho$ surface brightness profile is a consequence of imaging a dust volume density proportional to $r^{-2}$, where $r$ is the linear (3D) distance to the nucleus, i.e., a coma of long-lived dust produced by a constant radial outflow in the absence of acceleration by radiation pressure.  Using this model, we derive coma aperture corrections ranging from --0.42 to --0.11~mag for seeing values appropriate to our data.

\begin{figure*}
\includegraphics[width=0.7\textwidth]{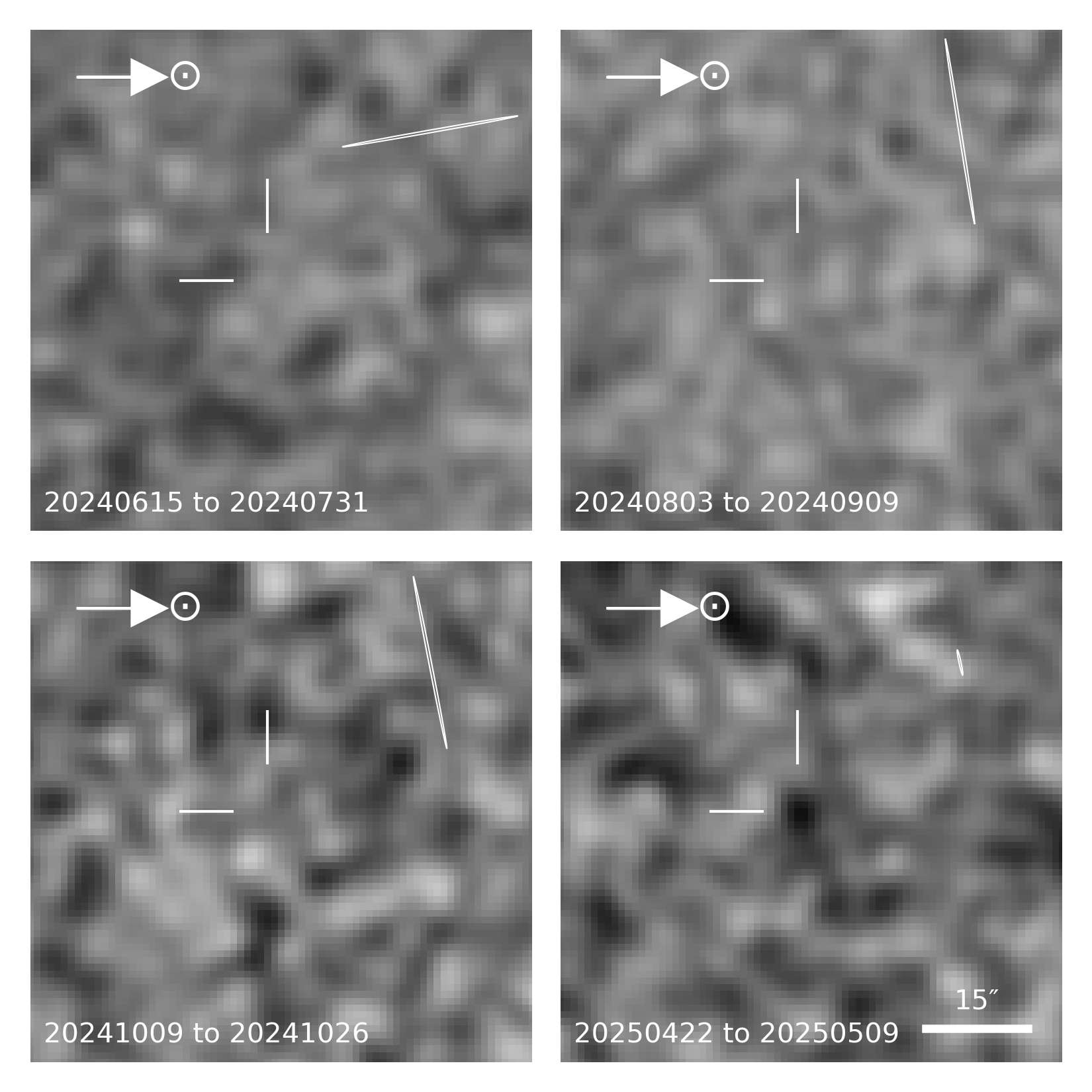}
\caption{Multi-night ZTF stacks centered at the nominal ephemeris position of 3I of periods from UT 2024 June 15 to UT 2025 May 9. Each stack is smoothed by a $2''$-wide Gaussian function for clarity. The $3\sigma$ uncertainty ellipses (appeared largely as a thin line due to very small semi-minor axis) are plotted in the upper-right corner of each panel. The images are color-inverted.}
\label{fig:medley_non}
\end{figure*}

3I was not visible on the nightly stacks before UT 2025 May 15. (There was a long data gap between UT 2024 November 21 and UT 2025 April 16 due to the comet's small solar elongation.) However, the comet was marginally visible at $1.6\sigma$ level at the ephemeris position when we combined all five non-detection nightly stacks in 2025, taken between 2025 April 22 and 2025 May 9 (Figure~\ref{fig:medley_non}, lower-right panel). We further validate this detection by combining the stacks in separate bands and confirming that the comet is still visible. To perform photometry on this multi-night, multi-band stack, we scaled the $g$-band stack to $r$-band using $g-r=0.65$~mag \citep[taking a spectral slope of 18\%/100~nm from][]{2025arXiv250705226O} in the PS1 photometric system, and obtained an aperture-corrected $r_\mathrm{PS1}=21.6\pm0.2$~mag. For the 2024 data, we split the data into three periods, each approximately $\sim1$~month long and containing 10--27 $g$- and $r$-band stacks. These stacks were then combined into three multi-night stacks, on which we performed color and aperture corrections as well as photometric measurements following the procedure described above. As shown in Figure~\ref{fig:medley_non}, 3I was not seen in any of these multi-nightly stacks. We derived the $3\sigma$ upper limits based on the standard deviation of 10,000 randomly placed apertures (3\arcsec{} radius) on each $5\arcmin\times5\arcmin$ cutout. In summary, we concluded that 3I was no brighter than a color-and-aperture-corrected $m_{\rm 3\sigma lim}\sim21$--$22$~mag on average over these periods.

We then fit the $r$-band data points using both the classic comet lightcurve model as well as the standard model for asteroids and inactive comets. The comet model is defined as

\begin{equation}
\label{eq:m1}
    m_1 = M_1 + 5 \log{\varDelta} + K_1 \log{r_\mathrm{H}} + \Phi(\alpha)
\end{equation}

\noindent where $m_1$ and $M_1$ are the apparent and absolute total magnitude of the comet, respectively, $\varDelta$ and $r_\mathrm{H}$ are geocentric and heliocentric distances in au, $K_1$ is the logarithmic heliocentric distance slope, and $\Phi(\alpha)$ is the phase function of the comet with respect to the phase angle $\alpha$, in which we use the Schleicher--Marcus phase function \citep[also referred to as the Halley--Marcus phase function;][]{2011AJ....141..177S}.

The ``bare nucleus'' model is a simplified version of Equation~\ref{eq:m1}:

\begin{equation}
\label{eq:nucleus}
    m_\mathrm{n} = H_\mathrm{n} + 5 \log{(\varDelta r_\mathrm{h})} + \Phi(\alpha)
\end{equation}

\noindent where $H_\mathrm{n}$ is the absolute magnitude of the cometary nucleus. For the phase function, we use a linear phase function $\Phi(\alpha)=\alpha \beta$ where $\beta=0.035~\mathrm{mag~deg^{-1}}$ is the phase coefficient of a typical comet \citep{2004come.book..223L}.

\begin{figure*}
\includegraphics[width=\textwidth]{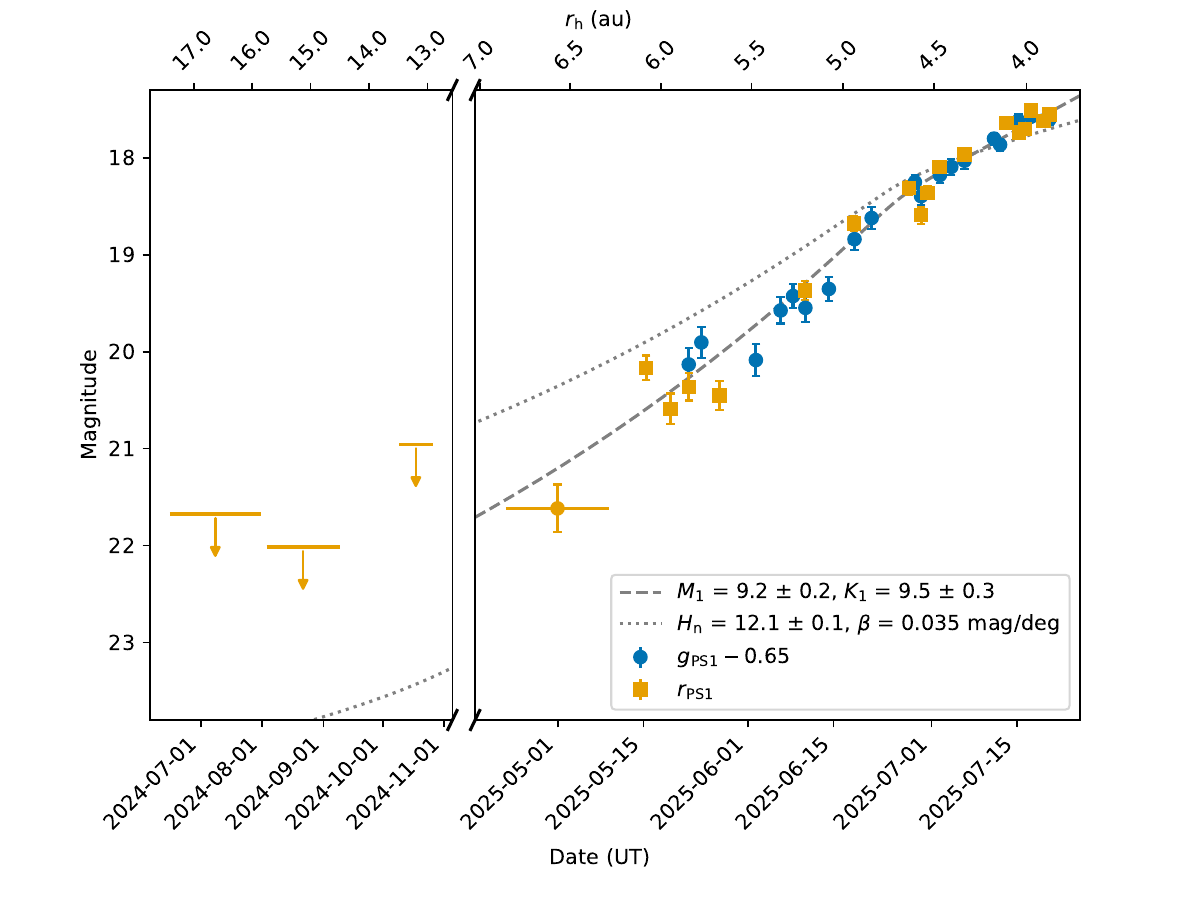}
\caption{Best-fit lightcurve of 3I using ZTF photometry from nightly and multi-nightly stacks from 2024 June 15 to 2025 July 20. Uncertainties of individual data points may be underestimated due to complications associated with crowded field photometry. For multi-nightly stacks, horizontal bars indicate the time bin sizes. The lightcurve models used are appropriate for a coma or nucleus, described by Eqs.~\ref{eq:m1} and \ref{eq:nucleus}, respectively.  For the 2024 panel, the coma model is fainter than 25.5 mag. We offset the $g$-band photometry by $-0.65$~mag using the spectral slope reported by \citet{2025arXiv250705226O}.}
\label{fig:lc}
\end{figure*}

Figure~\ref{fig:lc} shows the lightcurve as well as the uncertainty-weighted least-square fit. We derived $M_1=9.2\pm0.2$~mag, $K_1=9.5\pm0.3$, and $H_\mathrm{n}=H_{r,\mathrm{n}}=12.1\pm0.1$~mag.

\subsection{Dust Coma}\label{sec:dustanalysis}

These additional prediscovery detections allow us to peek into the early development of 3I's dust coma. To do so, we need to first isolate the signal from the cometary nucleus. \citet{jewitt2025_3IHST} estimated a lower limit absolute magnitude for 3I/ATLAS's nucleus of $H_{V,\mathrm{n}}>15.4$~mag from a convolutional surface brightness profile fitting analysis of high-resolution imaging from the \textit{Hubble Space Telescope (HST)}, which is the most stringent upper limit to the absolute nuclear magnitude reported to date.  Adopting this lower limit as the best currently available absolute nuclear magnitude, we can estimate the amount of excess ejected dust around the nucleus over the period covered by our observations and examine its evolution.

We first compute the expected apparent magnitude of the bare nucleus corresponding to each photometric point from ZTF using Equation~\ref{eq:nucleus} (with the same linear phase function with $\beta=0.035~\mathrm{mag~deg^{-1}}$) and the \textit{HST} constraint. We then subtract this value from the equivalent $V$-band magnitude for each measured photometric point transformed from $g_{\rm PS1}$ and $r_{\rm PS1}$ using a spectral slope of 18\%/100~nm as discussed in \S~\ref{sec:photometry}, to obtain $V$-band apparent magnitudes of the excess dust alone.  We then compute the equivalent $V$-band absolute magnitude of the dust, $M_{V,\mathrm{d}}$ (i.e., at $r_\mathrm{h}=\varDelta=1$~au and $\alpha=0^{\circ}$), again using Equation~\ref{eq:nucleus} but with the Schleicher--Marcus phase function described above for $\Phi(\alpha)$.  Finally, we compute the dust-to-nucleus flux ratios for each photometric point, $F_\mathrm{d}/F_\mathrm{n}$, using
\begin{equation}
    {F_\mathrm{d}\over F_\mathrm{n}} = 10^{0.4\left(H_{V,\mathrm{n}}-M_{V,\mathrm{d}}\right)}
\end{equation}

\begin{figure*}
\includegraphics[width=\textwidth]{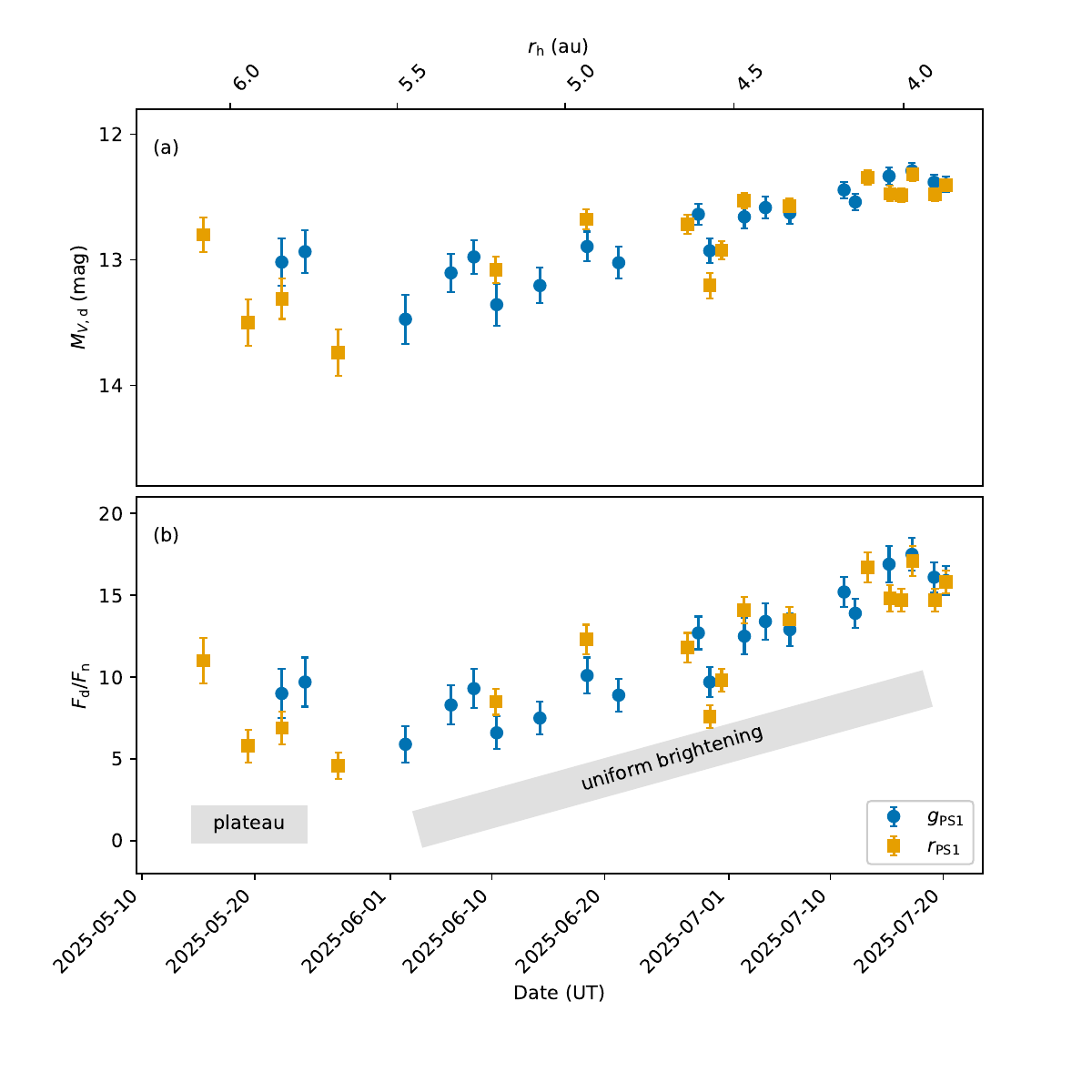}
\caption{Plots of (a) nucleus-subtracted $V$-band absolute magnitudes \citep[assuming a spectral slope of 18\%/100~nm;][]{2025arXiv250705226O} of 3I's excess dust inferred from the ZTF data, using the (lower limit) nuclear magnitude of $H_V>15.4$~mag reported by \citet[][dashed line]{jewitt2025_3IHST}, and (b) the dust-to-nucleus flux ratios calculated for each photometry point for 3I as functions of time (bottom axes) and heliocentric distance (top axes). The phases of ``plateau'' and ``uniform brightening'', described in the main text, are marked. Uncertainties of individual data points may be underestimated due to complications associated with crowded field photometry. 
}
\label{fig:dust_evolution}
\end{figure*}

Absolute magnitudes computed for excess dust and dust-to-nucleus flux ratios computed as described above are plotted as functions of time and heliocentric distance in Figure~\ref{fig:dust_evolution}.
We note a relatively uniform increasing/brightening trend in $F_\mathrm{d}/F_\mathrm{n}$ starting potentially as early as UT 2025 May 27 until the latest observations reported here on UT 2025 July 20.  Specifically, $F_d/F_n$ appears to increase approximately linearly from $F_\mathrm{d}/F_\mathrm{n}\sim5$ to $F_\mathrm{d}/F_\mathrm{n}\sim15$ in about 50 days.  This increasing trend appears to be preceded by a plateau of $F_\mathrm{d}/F_\mathrm{n}\sim5$ from the start of the detection on UT 2025 May 15 until May 24. The nature of this ``plateau'' is uncertain given the relatively larger error bars during this period, although we re-inspected the images and confirmed that these were not due to star contamination.

The ZTF dust-to-nucleus flux ratios were $\sim5\times$ larger than the ones found in an analysis of Rubin data obtained during the same period \citep[UT 2025 June 21 to July 2; see][\S~4.9.1]{2025arXiv250713409C}. This can be largely explained by the brighter nuclear magnitude used in the Rubin analysis ($H_{V,\mathrm{n}}=13.7$~mag compared to the \textit{HST}-derived $H_{V,\mathrm{n}}=15.4$~mag). If we used $M_{V,\mathrm{d}}=12.67$~mag as derived in the Rubin analysis \citep[][\S~4.9.2]{2025arXiv250713409C} and the \textit{HST}-derived $H_{V,\mathrm{n}}$, we obtain a dust-to-nucleus flux ratio of $\sim12$, which agrees well with the ZTF-derived number.

Assuming that the coma is optically thin (i.e., observed flux is directly proportional to total cross-sectional area), the observed flux of the coma can be used to estimate the dust production rate of the comet. Following the procedure described in \citet[][\S~2.3]{jewitt2025_3IHST}, we first calculate the cross-sectional area of the dust:

\begin{equation}
    C_\mathrm{d} = \frac{7.1\times10^{22}}{p_V} 10^{-0.4\left(M_{V,\mathrm{d}}-V_\odot\right)}
\end{equation}

\noindent where $p_V=0.04$ is the $V$-band geometric albedo of the dust grain, and $V_\odot=-26.7$ is the $V$-band magnitude of the Sun at 1~au. The dust production rate $\dot M_\mathrm{d}$ is then calculated by

\begin{equation}
    \dot M_\mathrm{d} = \frac{4 \rho_\mathrm{d} \bar a_\mathrm{d} C_\mathrm{d}}{3 s_\rho} \left( \frac{2\beta_\mathrm{d} g_1 X_{\mathrm{R},r_\mathrm{h}}}{r_\mathrm{h}^2} \right)^{0.5}
\end{equation}

\noindent where $\rho_\mathrm{d}$ is the bulk grain density of the dust grain, ${\bar a}_\mathrm{d}$ is the effective mean dust grain radius, $s_\rho$ is the radius of the photometric aperture projected at the comet's distance, $\beta_\mathrm{d} \sim 10^{-6}\cdot{\bar a}_\mathrm{d}^{-1}$ is the radiation pressure efficiency factor of grains at mean radius, $g_1=0.006~\mathrm{m~s^{-2}}$ is the solar gravitational acceleration at $r_{\rm h}=1$~au, and $X_{\mathrm{R},r_\mathrm{h}}=(5.9\times10^6)\cdot r_\mathrm{h}^{-0.5}~\mathrm{m}$ is the turning distance of dust grains in the sunward direction due to the deceleration by solar radiation, measured by \textit{HST} \citep{jewitt2025_3IHST} and scaled with $r_\mathrm{h}$. ($X_{\mathrm{R},r_\mathrm{h}}$ is most effectively constrained by \textit{HST}, as the angular distance of this turning point only amounts to $1.5''$ at $r_\mathrm{h}\sim4$~au, which is unresolvable for ZTF and most seeing-limited telescopes.)

Substituting the numbers we determined above and assuming ${\bar a}_\mathrm{d}=100$~\micron, we find $\dot M_\mathrm{d}\sim5~\mathrm{kg~s^{-1}}$ at the beginning of the observing window (early May 2025, $r_\mathrm{h}=6.0$~au) and $\dot M_\mathrm{d}\sim30~\mathrm{kg~s^{-1}}$ at the end of the window (mid July 2025, $r_\mathrm{h}=3.9$~au). The latter value corresponds to the conditions around the time of the Rubin and \textit{HST} observations (late June and mid July 2025) and agrees their reported rates within an order of magnitude\footnote{The original Rubin's ${\dot M}$ derived in \citet[][\S~4.9.2]{2025arXiv250713409C} is two orders of magnitude larger than the ZTF value; however, their Equation~6 was originally derived for \micron-sized particles. After accounting for the 100~\micron-sized particles being discussed here as well as the smaller nucleus constrained by the \textit{HST} data, we obtain ${\dot M}\sim60$~kg~s$^{-1}$.}. We note that our estimate does not consider the flux contribution from the nucleus, which constitutes no more than 20\% (Figure~\ref{fig:dust_evolution}b) of the total flux of the comet, and is therefore negligible for our order-of-magnitude level estimate.

\section{Discussion} \label{sec:disc}

The lightcurve in Figure~\ref{fig:lc} shows that the brightening of 3I inward of $r_\mathrm{h}=6.5$~au was consistent with an active comet. Specifically, 3I followed a brightening rate of $\propto r_\mathrm{h}^{-3.8}$ (note that $\propto r_\mathrm{h}^{-n}$ and $n=K_1/2.5$), which is steeper than 2I/Borisov \citep[$\propto r_\mathrm{h}^{-2.1}$;][]{2020AJ....159...77Y} -- the only other known interstellar comet to date, and is more akin to the dynamically old long-period comets and short-period comets in the Solar System \citep[$n\sim2.5$--$5.5$;][]{1978M&P....18..343W, 1995Icar..118..223A}, though we caution that most Solar System comets in those two papers were observed closer to the Sun. A more recent investigation by \citet[][specifically Figure~6 and \S~5.3]{2024PSJ.....5..273H} presented a slope out to $\sim9$~au, which had a modest sample size ($N=3$) but would suggest the same conclusion (i.e. a steeper brightening slope of dynamically old comets). 2I, on the other hand, is more in line with dynamically new comets. In the context of Solar System comets, a steeper brightening may indicate repeated heating of the comet, which could have occurred in its birth planetary system or during any prior encounters with other Solar System, coinciding with its likely old age \citep{2025arXiv250705318H} and thus possibly large number of stellar encounters \citep{2018ApJ...852L..13Z}. At the same time, impact gardening from the interstellar medium and cosmic ray bombardment could have also resurfaced the nucleus and exposed buried volatiles, potentially altering its response to solar heat \citep{stern86, 1996AJ....112.2310L}.

A weighted fit using the bare-nucleus lightcurve model yields $H_{r,\mathrm{n}}=12.1\pm0.1$~mag, but the fit deviates from the observational data over most of the period of observing. Early in the observing window, or 2025 April--May, the deviation reached $1.3$~mag, suggesting that the true $H_{r,\mathrm{n}}\geq13.4$~mag. This grossly agrees with the nuclear magnitude determined by Rubin \citep[$H_{r,\mathrm{n}}=13.2\pm0.2$;][]{2025arXiv250713409C}\footnote{Technically, the Rubin result is in $r_\mathrm{LSST}$ while ours is in $r_\mathrm{PS1}$, however they are virtually the same, as $r_\mathrm{LSST}-r_\mathrm{PS1}=-0.003$.}, although both are brighter than the \textit{HST}-estimated lower-limit absolute magnitude of $H_V>15.4$~mag as noted above. We conclude that the ``nucleus'' defined by both ZTF and Rubin corresponds to the material released in the earlier phase of cometary activity and is not the true nucleus.

Prediscovery photometry using archival ATLAS data dating back to 2025 March 8 was later reported by \citet{2025arXiv250905562T}, showing a brightening rate of $\propto r_\mathrm{h}^{-3.9}$ until the end of July 2025, consistent with our result. Specifically, they measured an \textit{o}-band magnitude of $o=21.6\pm0.2$ on 2025 April 27, equivalent to $r=22.1\pm0.2$, which also agrees our 2025 May 1 data point within uncertainty.

Another prediscovery photometry made by the \textit{Transiting Exoplanet Survey Satellite} back to 2025 May 7 were independently reported by \citet{2025arXiv250721967F} and \citet{2025arXiv250802499M}. They reported a mutually consistent nominal value of $T_\mathrm{mag}=20.9$~mag on 2025 May 7, with an uncertainty of 0.1--0.3~mag. Using the procedure described in \S~\ref{sec:analysis}, we derived an aperture correction of $-3.0$~mag assuming a square-shape \textit{TESS} aperture with 2-pixel ($42''$) sides. The color correction from \textit{TESS} $T_\mathrm{mag}$ to $r_\mathrm{PS1}$ was calculated using {\tt pysynphot} assuming solar color, which we derived $T_\mathrm{mag}-r_\mathrm{PS1}=-0.09$~mag. With both corrections applied, our 2025 May 1 detection of $r_\mathrm{PS1}=21.6\pm0.2$~mag translates to $T_\mathrm{mag}=18.6$~mag, which is $2.3$~magnitudes brighter than the actual \textit{TESS} detection. A straightforward explanation is that the coma no longer follows $1/\rho$ profile at \textit{TESS}'s pixel scale. Since $T_\mathrm{mag}$ is only $0.7$~mag brighter than the ZTF-measured brightness, we estimate that the $1/\rho$ profile only extends to $\sim5.7''$ from the nucleus. If we assume an expansion speed of $\sim10$~m/s, appropriate to 100~\micron-class dust grains at this distance from the Sun, we derive that the constant dust outflow started $\sim30$~days prior (in early April of 2025), at $r_\mathrm{h}=9$~au. We note that this coincides with the turn-on distance of CO$_2$ ice \citep{1979M&P....21..155C} as well as the activation of 2I \citep{2020AJ....159...77Y}.

The non-detection in 2024 showed that no strong outburst with $\Delta \mathrm{mag} \gtrsim 2$--$5$ (depends on the lightcurve model used) occurred during these periods, at $r_\mathrm{h}\sim13$--$17$~au. Based on this limited dataset, the lack of significant outburst appears to show that 3I is similar to 2I as well as most long-period Solar System comets \citep{2024PSJ.....5..273H}.

It is also worth noting that 3I was discovered at a larger heliocentric distance of $r_\mathrm{h}=4.5$~au compared to 2I, which was discovered relatively close to the Sun at $r_\mathrm{h}=3.0$~au. This difference is likely due to 3I's more favorable observing geometry in its inbound leg, rather than a difference in sizes or active levels of the comet, as both comets have a similar intrinsic brightness \citep[with absolute magnitudes $H\sim13$ at 5~au;][]{2020AJ....159...77Y}.

\section{Conclusion}

We present an analysis of ZTF prediscovery observations and constraints of 3I for its inbound leg up to $r_\mathrm{h}=17$~au, approximately a year before its discovery. We found that 3I had been active since at least $r_\mathrm{h}=6.5$~au. The comet followed a brightening rate of $\propto r_\mathrm{h}^{-3.8}$, which is significantly steeper than 2I and is more consistent with the dynamically old long-period comets and short-period comets in the Solar System. Analysis of the dust coma showed that 3I has a dust production rate of $\dot{M}\sim5~\mathrm{kg~s^{-1}}$ in 2025 May, increased to $\dot{M_\mathrm{d}}\sim30~\mathrm{kg~s^{-1}}$ towards mid-July 2025, appropriate to 100~\micron{} grains, which agrees with measurements by Rubin and \textit{HST} within the order of magnitude. Comparison with the prediscovery photometry derived from \textit{TESS} suggested that 3I started producing constant dust outflow probably as recently as 4 months before discovery, or $r_\mathrm{h}\sim9$~au, coinciding with the turn-on distance of CO$_2$ ice. We did not find signs of strong outbursts exhibited by 3I at $r_\mathrm{h}\sim13$--$17$~au from the Sun, a behavior that is consistent with most long-period Solar System comets as well as 2I.

\begin{acknowledgments}
QY is supported by NASA Grant 80NSSC21K0659. Based on observations obtained with the Samuel Oschin Telescope 48-inch at the Palomar Observatory as part of the Zwicky Transient Facility project. ZTF is supported by the National Science Foundation under Grants No.\ AST-1440341, AST-2034437, and currently Award \#2407588. ZTF receives additional funding from the ZTF partnership. Current members include Caltech, USA; Caltech/IPAC, USA; University of Maryland, USA; University of California, Berkeley, USA; University of Wisconsin at Milwaukee, USA; Cornell University, USA; Drexel University, USA; University of North Carolina at Chapel Hill, USA; Institute of Science and Technology, Austria; National Central University, Taiwan, and OKC, University of Stockholm, Sweden. Operations are conducted by Caltech's Optical Observatory (COO), Caltech/IPAC, and the University of Washington at Seattle, USA. 
\end{acknowledgments}





%
\facilities{PO:1.2m}

\software{astropy \citep{2013A&A...558A..33A, 2018AJ....156..123A, 2022ApJ...935..167A}, matplotlib \citep{hunter2007matplotlib}, numpy \citep{van2011numpy, harris2020array}, pysynphot \citep{2013ascl.soft03023S}, sbpy \citep{2019JOSS....4.1426M}}


\bibliography{sample701}{}
\bibliographystyle{aasjournalv7}



\end{CJK*}
\end{document}